# Optical response of metallic and insulating VO$_2$ calculated with the LDA approach

R.J.O. Mossanek and M. Abbate*

*Departamento de Física, Universidade Federal do Paraná,
Caixa Postal 19091, 81531-990 Curitiba PR, Brazil*

We calculated the optical response of metallic and insulating VO$_2$ using the LDA approach. The band structure calculation was based in the full-potential linear-muffin-tin method. The imaginary part of the dielectric function $\varepsilon_2(\omega)$ is related to the different optical transitions. The Drude tail in the calculation of the metallic phase corresponds to intra-band d-d transitions. The calculation in the insulating phase is characterized by the transitions to the $d_\parallel^*$ band. The low frequency features, 0.0 – 5.0 eV, correspond to V 3d – V 3d transitions, whereas the high frequency structures, 5.0 – 12 eV, are related to O 2p – V 3d transitions. The calculation helps to explain the imaginary part of the dielectric function $\varepsilon_2(\omega)$, as well as the electron-energy-loss and reflectance spectra. The results reproduce not only the energy position and relative intensity of the features in the spectra, but also the main changes across the metal-insulator transition and the polarization dependence. The main difference is a shift of about 0.6 eV in the calculation of the insulating phase. This discrepancy arises because the LDA calculation underestimates the value of the band gap.

## I. Introduction

Several early transition metal oxides exhibit interesting metal-insulator (MIT) transitions [1]. For example, the VO$_2$ compound presents a first order metal-insulator transition around $T_C = 340$ K [2,3]. Below the transition temperature VO$_2$ is a diamagnetic insulator with a monoclinic structure, whereas above the critical temperature it becomes a paramagnetic metal with a tetragonal structure. The insulating structure is characterized by the formation of distinct V-V dimmers. The V 3d levels are split by crystal field effects into three lower $t_{2g}$ and two higher $e_g$ levels. The $t_{2g}$ level which connects the V ions along the c-axis forms the so-called $d_\parallel$ band. The V-V dimmerization in the insulating phase splits the $d_\parallel$ band opening a band gap [4].

The electronic structure of VO$_2$ was investigated by photoemission [5,6] and X-ray absorption [7,8]. The orbital occupancy across the transition was determined using polarized X-ray absorption [9]. The electronic structure of VO$_2$ was calculated using the LDA [10,11] and the cluster model methods [12]. The charge-transfer satellites in the core-level spectra were analyzed using the cluster model approach [13,14]. Recent photoemission studies showed strong correlation effects in the V 3d region of the spectra [15,16]. These effects were studied using the GW [17], the DMFT [18], and the cluster model [19] methods. The microscopic origin of the MIT transition in VO$_2$ was also studied using the DMFT method [20,21].

The optical properties of VO$_2$ were studied using the reflectivity and transmission methods [22,23]. More recent studies included detailed reflectivity, EELS and optical conductivity measurements [24-26]. However, the optical response of VO$_2$ was calculated only in the insulating phase [17]. The Drude tail in the metallic phase was assigned to d–d transitions across the Fermi level. There is a discrepancy in the identification of the additional optical transitions above 1.5 eV. These features were attributed to charge transfer [24], Mott-Hubbard [26], and quasi-particle transitions [27].

We studied here the optical response of both metallic and insulating VO$_2$ using the LDA approach. The structures in the dielectric function $\varepsilon(\omega)$ are related to the optical transitions in the density of states. The imaginary part of the dielectric function $\varepsilon_2(\omega)$ is in relatively good agreement with the experiment. The main discrepancy in the calculation is a rigid shift of about 0.6 eV in the insulating phase, because the LDA calculation tends to underestimate the value of the band gap in this phase. The calculation shows that the features in the spectra up to 5.0 eV are mostly related to d–d transitions. Further, the calculated reflectivity and EELS spectra are also in good agreement with the experiment.

## II. Calculation Details

The band structure was calculated using the Full-Potential Linear-Muffin-Tin Orbital method [21]. The exchange and correlation potential was determined using the Vosko approximation. The imaginary part of the dielectric function $\varepsilon_2(\omega)$ is calculated from the optical transitions, whereas the corresponding real part $\varepsilon_1(\omega)$ is obtained using the Kramers-Kronig transformation [22]. The EELS spectrum $L(\omega)$ was obtained from the dielectric function using the following relation:

$$L(\omega) = \text{Im}\left[-\frac{1}{\varepsilon(\omega)}\right] \quad (1)$$

The reflectivity coefficient $R(\omega)$ at normal incidence was calculated using the Fresnel equations:

$$R(\omega) = \left|\frac{\sqrt{\varepsilon(\omega)}-1}{\sqrt{\varepsilon(\omega)}+1}\right|^2 \quad (2)$$

The calculation results were convoluted with a Gaussian function with a FWHM of 0.5 eV.

The metallic phase was calculated in the tetragonal structure (space group P4_2/mnm). The lattice parameters were a = 4.530 Å and c = 2.896 Å. The atomic positions were V = (0.000, 0.000, 0.000) and O = (0.305, 0.305, 0.000). The self-consistent potential, the density of states, and dielectric function were calculated using 24 irreducible **k**-points.

The insulating phase was calculated in the monoclinic structure (space group P2_1/c). The lattice parameters were a = 5.743 Å, b = 4.517 Å, c = 5.375 Å and β = 121.56°. The atomic positions were V = (0.233, 0.024, 0.021), O1 = (-0.118, 0.288, 0.272) and O2 = (0.399, 0.315, 0.293). The self-consistent potential, the density of states, and the dielectric function were calculated using 80 irreducible **k**-points.

## III. Results

The upper part of Fig. 1 shows the calculated density of states of $VO_2$ in the metallic phase. The calculation corresponds to a metallic solution with a continuous DOS at the Fermi level. The DOS is composed by the O 2p band, from –8.0 to –2.0 eV; and the V 3d band, from –0.5 to 5.0 eV. The V 3d band is split by crystal field effects into the $t_{2g}$, from -0.5 to 2.0 eV, and the $e_g$ bands, from 2.0 to 5.0 eV. The contribution of the $d_\parallel$ band in the metallic phase is mostly concentrated close to the Fermi level. The arrows indicate the main optical transitions across the Fermi level in the metallic phase. They consist of the d-d transitions, from the occupied $t_{2g}$ to the unoccupied $t_{2g}$ (A) and $e_g$ states (B), as well as the p-d transitions, from non-bonding O 2p states to unoccupied $t_{2g}$ (C) and $e_g$ states (D).

Figure 2 shows the imaginary part $\varepsilon_2(\omega)$ and the real part $\varepsilon_1(\omega)$ of the dielectric function of metallic $VO_2$. The solid (dotted) line corresponds to polarization parallel (perpendicular) to the rutile c-axis. The imaginary part $\varepsilon_2(\omega)$ presents the characteristic Drude tail A at low frequencies, which is attributed to intra-band d-d transitions from the occupied $t_{2g}$ to the unoccupied $t_{2g}$ states. The next feature B is ascribed to d-d transitions from the occupied $t_{2g}$ to the unoccupied $e_g$ states. Finally, the structures C and D correspond to p-d transitions to $t_{2g}$ and $e_g$ states, respectively. These transitions come from mostly non-bonding O 2p states at the top of the valence band. Finally, the polarization dependence of the dielectric function in this phase is relatively weak.

The O 2p states in the valence band are split by O 2p – O 2p and O 2p – V 3d interactions [19]. The $E_g$ and $T_{2g}$ symmetries are related to a covalent mixture of O 2p and V 3d states, whereas the $A_{1g}$, $T_{1g}$, $2T_{1u}$, and $T_{2u}$ symmetries correspond to non-bonding O 2p states. Due to the dipole selection rule, only the $T_{1u}$ symmetry is allowed to reach the V 3d states. Thus the p-d transitions described above comes from a non-bonding O 2p state with a $T_{1u}$ symmetry. According to a recent cluster model calculation, the first state with $T_{1u}$ symmetry appears about 4.0 eV. The energy of this $T_{1u}$ state is in good agreement with the origin of the p-d transitions depicted in Fig. 1.

Figure 3 compares the calculated $\varepsilon_2(\omega)$ of metallic $VO_2$ with the experimental results taken from Ref. 22. The calculation reproduces the energy positions and the relative intensity of the different features. The weak polarization dependence of the experimental results is also reproduced by the calculation. The Drude tail A up to 1.5 eV corresponds to intra-band d-d transitions to unoccupied $t_{2g}$ states, whereas the structure B from 2.0 to 5.0 eV is due to d-d transitions to unoccupied $e_g$ states. The lower panel of Fig. 3 shows the total unoccupied density of states of metallic $VO_2$ for comparison. The resemblance with the DOS confirms that the structures in $\varepsilon_2(\omega)$ are indeed due to d-d transitions.

The lower part of Fig. 1 shows the calculated density of states of $VO_2$ in the insulating phase. The calculation corresponds to a semi-metallic solution with a pseudo-gap of 0.2 eV at the Fermi level. The LDA calculation underestimates the experimental value of the band gap of about 0.7 – 0.9 eV. The DOS is composed by the O 2p band, from –7.0 to –1.0 eV, and the V 3d band, from –0.5 to 5.0 eV. The V 3d band is again split by crystal field effects into the $t_{2g}$, from -0.5 to 2.3 eV, and the $e_g$ bands, from 2.3 to 5.0 eV. The $d_\parallel$ band in the insulating phase is split due to bonding interactions within the V–V dimers [4]. The bonding part ($d_\parallel$) appears just below the Fermi level, whereas the anti-bonding part ($d_\parallel^*$) is shifted to around 2.2 eV. The arrows represent again the different optical transitions across the Fermi level in the insulating phase. They consist of a distinct d-d transition, from the occupied $d_\parallel$ to the unoccupied

$d_\parallel^*$ bands (A'), additional d-d transitions, from the occupied $d_\parallel$ to the unoccupied $t_{2g}$ (A) and $e_g$ states (B), and finally the p-d transitions, from non-bonding O 2p states to the unoccupied $t_{2g}$ (C) and $e_g$ states (D).

Figure 4 shows the imaginary part $\varepsilon_2(\omega)$ and the real part $\varepsilon_1(\omega)$ of the dielectric function of insulating $VO_2$. The solid (dotted) line corresponds to polarization parallel (perpendicular) to the pseudo-rutile c-axis. The imaginary part $\varepsilon_2(\omega)$ shows that the Drude tail disappears in the insulating phase. The first structure A is assigned to d-d transitions from the occupied $d_\parallel$ band to the unoccupied $t_{2g}$ states, the peak A' is attributed to d-d transitions to the unoccupied $d_\parallel^*$ band, and the feature B is ascribed to d-d transitions to empty $e_g$ states. The assignment of the A' structure is confirmed by its strong polarization along the parallel direction. Finally, the structures C and D correspond to p-d transitions to unoccupied $t_{2g}$ and $e_g$ states, respectively. These transitions come again from mostly non-bonding O 2p states at the top of the valence band. The polarization dependence of the dielectric function is much larger in the insulating phase. The angular dependence of the peak A', which is related to $d_\parallel$-$d_\parallel^*$ transitions, is particularly strong.

Figure 5 compares the calculated $\varepsilon_2(\omega)$ of insulating $VO_2$ with the experimental results taken from Ref. 22. The calculation was rigidly shifted by 0.6 eV to take into account the underestimated LDA band gap. The calculated $\varepsilon_2(\omega)$ reproduces reasonably well the energy positions of the different features. But the intensity of the A and A' structures, as well as their angular dependence, is overestimated. The peak A corresponds to d-d transitions to unoccupied $t_{2g}$ states, the peak A' is due to d-d transitions to the unoccupied $d_\parallel^*$ band, and the structure B is related to d-d transitions to unoccupied $e_g$ states. The lower panel of Fig. 5 shows the total unoccupied density of states of insulating $VO_2$. The close similarity with the DOS corroborates again that the structures in $\varepsilon_2(\omega)$ are due to d-d transitions.

Figure 6 compares the calculated EELS spectra of $VO_2$ with the experimental results taken from Ref. 25. The upper panel corresponds to the metallic phase, whereas the lower panel gives the insulating phase results. The insulating phase calculation was shifted 0.6 eV to compensate the underestimated LDA band gap. The calculation reproduces the energy position and relative intensity of the structures in the experiment. In addition, the calculation explains the changes across the MIT transition over a broad frequency range. The maxima in the loss function $L(\omega)$ correspond to a combined minima in both $\varepsilon_1(\omega)$ and $\varepsilon_2(\omega)$. The peak around 1.5 eV in the metallic phase is related to the minimum of both $\varepsilon_1(\omega)$ and $\varepsilon_2(\omega)$, see Fig. 2. This peak decreases in insulating $VO_2$ because the $d_\parallel^*$ bands is shifted to this energy region, reducing the extent of the minimum in $\varepsilon_2(\omega)$ and producing a relative maximum in $\varepsilon_1(\omega)$, see Fig. 4. The minima in both $\varepsilon_1(\omega)$ and $\varepsilon_2(\omega)$ also explain the bumps about 4–5 eV and the peaks around 10–12 eV.

Figure 7 compares the calculated reflectance of $VO_2$ with the experimental results taken from Ref. 24. The upper panel gives the metallic phase results, whereas the lower panel corresponds to the insulating phase. The insulating phase calculation was shifted 0.6 eV to compensate the underestimated LDA band gap. The metallic calculation reproduces reasonably well the energy positions and intensity of the main features. Even the relatively weak polarization dependence of the structures is reflected in the calculation. The features in the reflectance spectra can be related to the different structures in the dielectric functions. The Drude tail A and the feature B are related to d-d transitions, whereas the structures C and D corresponds to p-d transitions. The insulating phase calculation is again in relatively good agreement with the experiment. Even the polarization dependence of the spectra is qualitatively reproduced by the calculation. The structures A/A' and B are related to d-d transitions, whereas the features C and D corresponds to p-d transitions.

The LDA approach provides a good starting point to analyze the optical properties of $VO_2$. The main discrepancy is a rigid shift in the insulating phase because LDA underestimate the band gap. Despite this drawback, the LDA calculation helps to explain the optical response of $VO_2$. The experiment analyzed includes the imaginary part of the dielectric function $\varepsilon_2(\omega)$, as well as the electron-energy-loss and the reflectance spectra. The results reproduce not only the position and relative intensity of the features in the spectra, but also the main changes across the MIT transition and the relative polarization dependence. The agreement is reasonably good over a relatively broad, 0.5-12 eV, energy range. This would suggest that excitonic effects do not play a dominant role in this system (the same conclusion was obtained in the previous GW calculation of insulating $VO_2$ [17]). The Drude tail in the calculation of the metallic phase corresponds to intra-band d-d transitions. The calculation in the insulating phase is characterized by the transitions to the $d_\parallel^*$ band. The above results indicate that the response from 0.5 to 5.0 eV is due to d-d transitions, whereas the features from 5.0 to 12.0 eV correspond to p-d transitions.

## IV. Summary and conclusions

In summary, we calculated the optical response of metallic and insulating $VO_2$ using the LDA approach. The calculation results were compared to the experimental $\varepsilon_2(\omega)$, EELS and reflectivity spectra. The results reproduce not only the position and relative intensity of the features in the spectra, but also the main changes across the MIT transition and the relative polarization dependence. The LDA approach thus provides a good starting point to analyze the optical properties of $VO_2$. The main discrepancy is a rigid shift in the insulating phase because LDA underestimate the band gap. The Drude tail in the calculation of the metallic phase corresponds to intra-band d-d transitions. The calculation in the insulating phase is characterized by the transitions to the $d_{\parallel}$* band. The low frequency features in the optical spectra, from 0.5 to 5.0 eV, are related to d-d transitions, whereas the high energy structures, from 5.0 to 12.0 eV, are attributed to p-d transition.

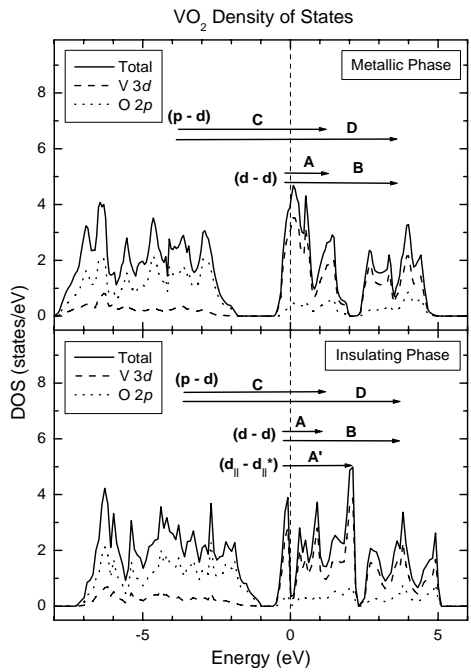

**Figure 1:** Calculated density of states of metallic and insulating VO$_2$. The arrows represent the transitions assigned to the features in the optical spectra.

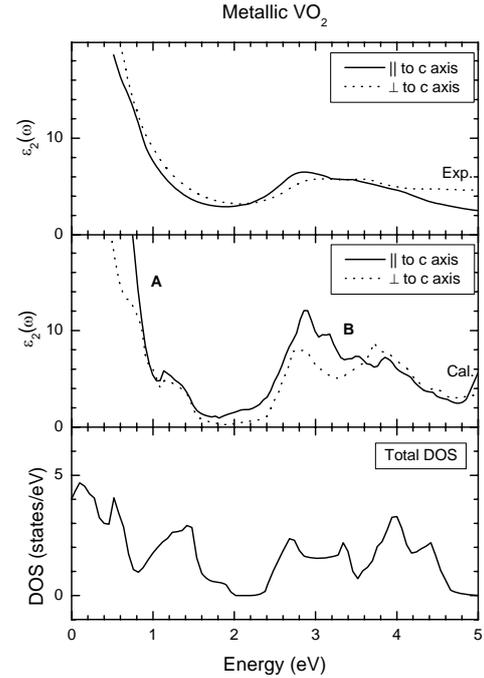

**Figure 3:** Imaginary part of the dielectric function $\varepsilon_2(\omega)$ of metallic VO$_2$ compared to the unoccupied DOS and experimental results taken from Ref. 22.

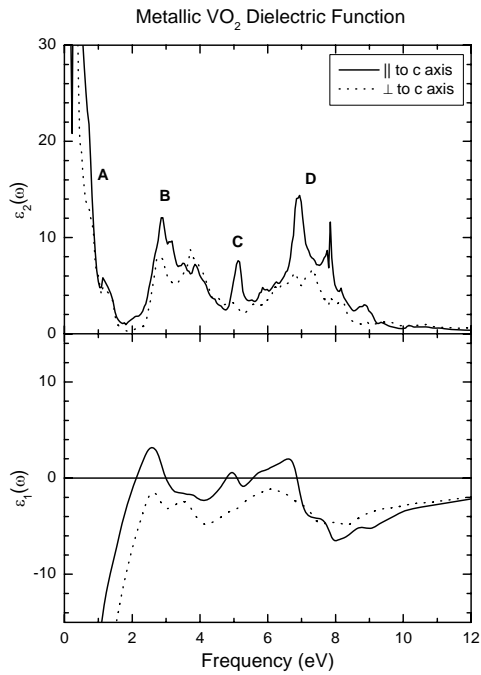

**Figure 2:** Calculated dielectric function of metallic VO$_2$ decomposed in parallel (solid line) and perpendicular (dotted line) to the c-axis.

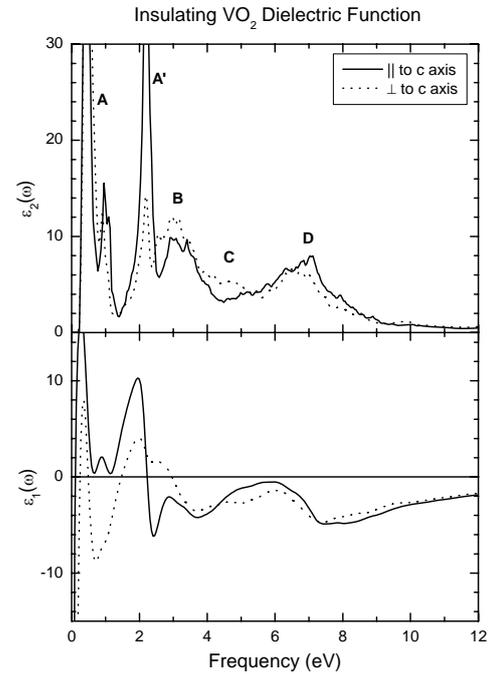

**Figure 4:** Calculated dielectric function of insulating VO$_2$ decomposed in parallel (solid line) and perpendicular (dotted line) to the c-axis.

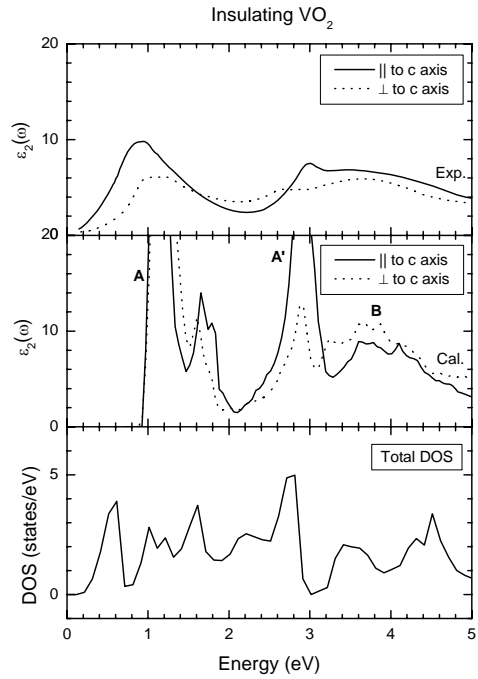

**Figure 5:** Imaginary part of the dielectric function $\varepsilon_2(\omega)$ of insulating $VO_2$ compared to the unoccupied DOS and experimental results taken from Ref. 22.

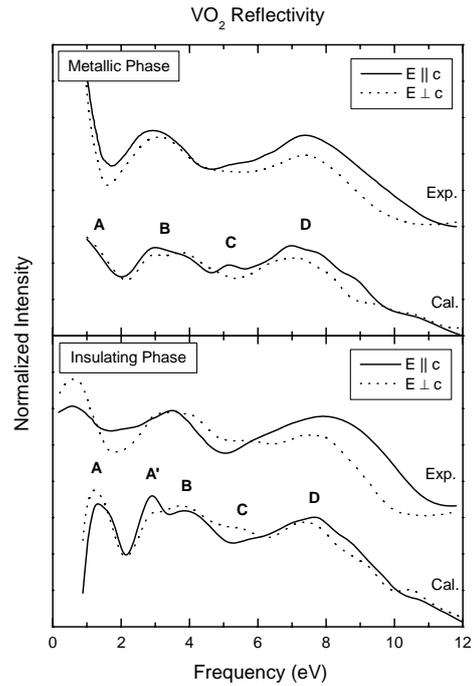

**Figure 7:** The calculated reflectivity spectra of metallic and insulating $VO_2$, decomposed in the parallel (solid line) and perpendicular (dotted line) to c axis components, compared to experimental results taken from Ref. 24.

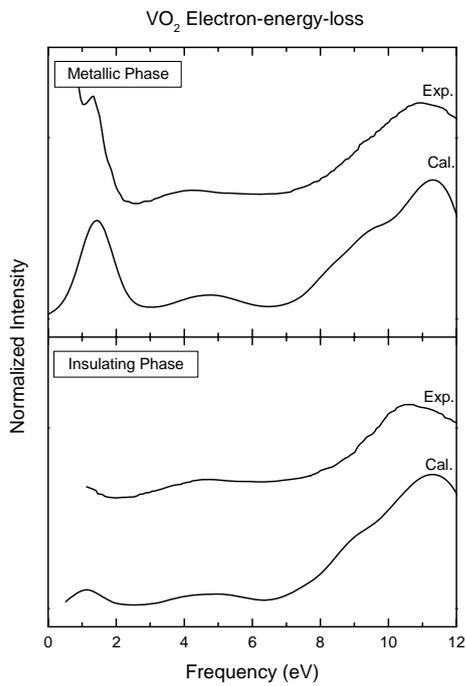

**Figure 6:** Calculated electron-energy-loss spectra of metallic and insulating $VO_2$ compared to experimental results taken from Ref. 25.